\documentclass[paper]{JHEP3}

\usepackage{cite}
\usepackage{epsfig}
\usepackage{rotating}

\newcommand{\beq}{\begin{equation}}
\newcommand{\eeq}{\end{equation}}
\newcommand{\bqa}{\begin{eqnarray}}
\newcommand{\eqa}{\end{eqnarray}}
\newcommand{\nl}{\nonumber \\}
\def\db#1{\bar D_{#1}}
\def\zb#1{\bar Z_{#1}}
\def\d#1{D_{#1}}
\def\tld#1{\tilde {#1}}

\def\eqn#1{Eq.~(\ref{#1})}

\def\eqnss#1#2{Eqs.~(\ref{#1})-(\ref{#2})}
\def\fig#1{Fig.~{\ref{#1}}}

\def\app#1{Appendix~\ref{#1}}

\newcommand\fverb{\setbox\pippobox=\hbox\bgroup\verb}
\newcommand\fverbdo{\egroup\medskip\noindent%
                        \fbox{\unhbox\pippobox}\ }
\newcommand\fverbit{\egroup\item[\fbox{\unhbox\pippobox}]}
\newbox\pippobox

\def\spa#1.#2{\left\langle#1\,#2\right\rangle}
\def\spb#1.#2{\left[#1\,#2\right]}

\def\feynsl#1{
  \setbox0=\hbox{/} \setbox1=\hbox{$#1$}
  \dimen0=\wd0 \advance\dimen0 by -\wd1 \divide\dimen0 by 2
  \ifdim\wd0>\wd1 \raise.15ex\copy0\kern-\wd0\kern\dimen0\copy1\kern\dimen0
  \else \kern-\dimen0\raise.15ex\copy0\kern-\dimen0\kern-\wd1\copy1\fi}

\newskip\humongous \humongous=0pt plus 1000pt minus 100pt

\newif\ifdtup

\def    \br(#1,#2)          {\mbox{$\langle #1 \, #2 \rangle$}}
\def    \sq(#1,#2)          {\mbox{$\left[  #1 \, #2 \right]$}}

\title{Numerical Evaluation of Six-Photon Amplitudes
}

\author{Giovanni Ossola\footnote{
e-mail: ossola@inp.demokritos.gr} , Costas G.
Papadopoulos\footnote{e-mail: costas.papadopoulos@cern.ch}
\\ Institute
of Nuclear Physics, NCSR "DEMOKRITOS", 15310 Athens, Greece.}

\author{Roberto Pittau\footnote{
e-mail: roberto.pittau@to.infn.it}
\\ Dipartimento di Fisica
Teorica, Univ. di Torino and INFN, sez. di Torino,
Italy.}

\abstract{We apply the recently proposed amplitude reduction at the
integrand level method, to the computation of the scattering process
$2 \gamma \to 4 \gamma$, including the case of a massive fermion
loop. We also present several improvements of the method, including
a general strategy to reconstruct the rational part of any one-loop
amplitude and the treatment of vanishing Gram-determinants.}
\preprint{}


\begin{document}

\section{Introduction}
In the last few years a big effort has been devoted by several authors to the
problem of an efficient computation of one-loop corrections for
multi-particle processes. This is a problem relevant for both  LHC and ILC
physics. In the case of QCD, the NLO six gluon amplitude
has been recently obtained by two different groups 
\cite{Ellis:2006ss}, and, in the case of $e^+ e^-$ collisions, 
complete EW calculations, involving 5-point
\cite{Belanger:2003ya} and 6-point \cite{Denner:2005fg} loop functions
are available at the cross section level.
 The used techniques range from purely numerical methods to
analytic ones, also including semi-numerical approaches. For analytical
approaches, the main issue is reducing, using computer algebra,
generic one-loop integrals into a minimal set of scalar integrals
(and remaining pieces, the so called rational terms), mainly by
tensor 
reduction~\cite{'tHooft:1978xw,Passarino:1978jh,Denner:2005nn,
Binoth:1999sp}. For
multi-particle processes though this method becomes quite cumbersome
because of the large number of terms generated and the appearance of
numerical instabilities due to the zeros of Gram-determinants. On
the other hand, several numerical or semi-numerical methods aim for
a direct numerical computation of the tensor
integrals~\cite{numerical}. Although purely numerical methods can in
principle deal with any configuration of masses and also allow for a
direct computation of both non-rational and rational terms, their
applicability remains limited due to the high demand of
computational resources and the non-existence of an efficient
automation.

In a different approach, the one-loop amplitude rather than
individual integrals are evaluated using the unitarity cut
method~\cite{unitarity-cut}, which relies on tree amplitudes and
avoids the computation of Feynman diagrams. In another development,
the four-dimensional unitarity cut method has been used for the
calculation of QCD amplitudes~\cite{Britto:2004nc}, using
twistor-based approaches~\cite{twistors}. Moreover, a generalization
of the the unitarity cut method in $d$ dimensions, has been pursued
recently~\cite{Anastasiou}.

Nevertheless, in practice, only the part of the amplitude
proportional to the loop scalar functions can be obtained
straightforwardly. The remaining piece, the rational part, should
then be reconstructed either by using a direct computation based on
Feynman diagrams~\cite{Binoth1,Binoth2,Xiao} or by using a bootstrap
approach~\cite{Bern:2005cq}. Furthermore the complexity of the
calculation increases away from massless theories.

In a recent paper \cite{Ossola:2006us}, we proposed a reduction
technique for arbitrary one-loop sub-amplitudes at {\it the
integrand level} by exploiting numerically the set of kinematical
equations for the integration momentum, that extend the quadruple,
triple and double cuts used in the unitarity-cut method. The method
requires a minimal information about the form of the one-loop
(sub-)amplitude and therefore it is well suited for a numerical
implementation. The method works for any set of internal and/or
external masses, so that one is able to study the full electroweak
model, without being limited to massless theories.

In this paper, we describe our experience with the first practical
non-trivial implementation of such a method in the computation of a
physical process: namely $2 \gamma \to 4 \gamma$, including
massive fermion loops.
For the massless case, there are a few results available in the literature.
Analytical expressions were first presented by Mahlon \cite{mahlon} some time
ago, however his results do not cover all possible helicity configurations.
More recently the complete set of six-photon amplitudes was computed
numerically by Nagy and Soper \cite{nagy}. Very recently the same results were
also obtained by Binoth et al. \cite{sixph}, that also provide compact
analytical expressions.

In section \ref{ratpart}, we recall the basics of our method
and, in particular, we show how the knowledge of the rational terms
can be inferred, with full generality, once the coefficients of the
loop functions have been determined.

 In section \ref{numinacc}, we outline our solution to cure the numerical
inaccuracies related to the appearance of zeros of Gram-determinants. 
We explicitly illustrate the case of 2-point amplitudes,
that we had to implement to deal with the process at hand.

 In section \ref{numerics}, we present our numerical results. For massless fermion loops
we compare with available results. Moreover, since we are not
limited to massless contributions, we also present, for the first
time, results with massive fermion loops.

Finally, in the last section, we discuss our conclusions and future
applications.

\section{The method and the computation of the rational terms \label{ratpart}}
The starting point of the method is the general expression for the
{\it integrand} of a generic $m$-point one-loop (sub-)amplitude
\cite{Ossola:2006us}
\bqa
\label{eq:1}
A(\bar q)= \frac{N(q)}{\db{0}\db{1}\cdots \db{m-1}}\,,~~~
\db{i} = ({\bar q} + p_i)^2-m_i^2\,,~~~ p_0 \ne 0\,,
\eqa
where we use a bar to denote objects living
in $n=~4+\epsilon$  dimensions, and $\bar q^2= q^2+ \tld{q}^2$
\footnote{$\tld{q}^2$ is $\epsilon$-dimensional and $(\tld{q} \cdot q) = 0$.}.
In the previous equation, $N(q)$ is the 4-dimensional part of the
numerator function of the amplitude \footnote{If needed,
the $\epsilon$-dimensional part of the numerator should be treated 
separately, as explained in \cite{Pittausimple}.}.
$N(q)$ depends on the $4$-dimensional denominators
$\d{i} = ({q} + p_i)^2-m_i^2$ as follows
\bqa
\label{eq:2}
N(q) &=&
\sum_{i_0 < i_1 < i_2 < i_3}^{m-1}
\left[
          d( i_0 i_1 i_2 i_3 ) +
     \tld{d}(q;i_0 i_1 i_2 i_3)
\right]
\prod_{i \ne i_0, i_1, i_2, i_3}^{m-1} \d{i} \nl
     &+&
\sum_{i_0 < i_1 < i_2 }^{m-1}
\left[
          c( i_0 i_1 i_2) +
     \tld{c}(q;i_0 i_1 i_2)
\right]
\prod_{i \ne i_0, i_1, i_2}^{m-1} \d{i} \nl
     &+&
\sum_{i_0 < i_1 }^{m-1}
\left[
          b(i_0 i_1) +
     \tld{b}(q;i_0 i_1)
\right]
\prod_{i \ne i_0, i_1}^{m-1} \d{i} \nl
     &+&
\sum_{i_0}^{m-1}
\left[
          a(i_0) +
     \tld{a}(q;i_0)
\right]
\prod_{i \ne i_0}^{m-1} \d{i} \nl
     &+& \tld{P}(q)
\prod_{i}^{m-1} \d{i}\,. \eqa
Inserted back in \eqn{eq:1}, this expression
simply states the multi-pole nature of any $m$-point one-loop amplitude,
that, clearly, contains a pole for any
propagator in the loop, thus one has terms ranging from 1 to $m$ poles.
 Notice that the term with no poles, namely that one proportional to
$\tld{P}(q)$ is polynomial and vanishes upon integration
in dimensional regularization; therefore does not contribute to the amplitude,
as it should be.
  The coefficients of the poles can be further split in two pieces.
A piece that still depend on $q$ (the terms
$\tld{d},\tld{c},\tld{b},\tld{a}$), that vanishes upon integration,
and a piece that do not depend on q (the terms $d,c,b,a$).
 Such a separation is always possible, as shown in Ref.~\cite{Ossola:2006us}, and, with
this choice, the latter set of coefficients is therefore immediately
interpretable as the ensemble of the 
coefficients of all possible 4, 3, 2, 1-point
one-loop functions contributing to the amplitude.

 Once \eqn{eq:2} is established, the task of computing the one-loop amplitude
is then reduced to the algebraical problem of determining
the coefficients $d,c,b,a$ by evaluating the function $N(q)$
a sufficient number of times, at different values of $q$,
and then inverting the system.
That can be achieved quite efficiently by singling out
particular choices of $q$ such that, systematically,
4, 3, 2 or 1 among all possible denominators $\d{i}$ vanishes.
 Then the system of equations is solved iteratively.
First one determines all possible 4-point functions,
then the 3-point functions and so on.
 For example, calling $q_0^\pm$ the 2 (in general complex) solutions for which
\bqa \d{0}= \d{1}= \d{2}=\d{3} = 0\,, \eqa (there are 2 solutions because
of the quadratic nature of the propagators) and since the functional
form of $\tld{d}(q;0123)$ is known, one directly finds the coefficient
of the box diagram containing the above 4 denominators through
the two simple equations
\bqa
N(q_0^\pm) &=& [d(0123) + \tld{d}(q_0^\pm;0123)] \prod_{i\ne 0,1,2,3}
\d{i} (q_0^\pm)
\,.
\eqa
This algorithm also works in the case of 
complex denominators, namely with complex masses.
 Notice that the described procedure can be performed
{\em at the amplitude level}. One does not need to
repeat the work for all Feynman diagrams, provided their sum is known:
we just suppose to be able to compute $N(q)$ numerically.

 As a further point notice that, since the terms
$\tld{d},\tld{c},\tld{b},\tld{a}$ still depend on $q$, also the
separation among terms in \eqn{eq:2} is somehow arbitrary.
 Terms containing a different numbers of denominators
can be shifted from one piece to the other in \eqn{eq:2},
by relaxing the requirement that the integral over
the terms containing $q$ vanishes. This fact provides an handle to cure
numerical instabilities occurring at exceptional phase-space points.
In Section \ref{numinacc} we will show in detail such a mechanism at work
for the 2-point part of the amplitude.

 The described procedure works without any modification in 4 dimensions.
However, even when starting from a perfectly finite tensor integral,
the tensor reduction may eventually lead to integrals
that need to be regularized.
 A typical example are the rank six 6-point functions contributing
to the scattering $2 \gamma \to 4 \gamma$ we want to study.
 Such tensors are finite, but tensor reduction iteratively leads to
rank $m$ $m$-point tensors with $ 1 \le m \le 5 $, that are
ultraviolet divergent when $m \le 4$.
For this reason, we introduced, in \eqn{eq:1}, the $d$-dimensional
denominators $\db{i}$, that differs by an amount $\tld{q}^2$ from
their 4-dimensional counterparts
\bqa
\db{i}= \d{i} + \tld{q}^2\,.
\eqa
The result of this is a mismatch in the cancellation
of the $d$-dimensional denominators of \eqn{eq:1} with the $4$-dimensional
ones of \eqn{eq:2}. The rational part of the amplitude comes from
such a lack of cancellation.

In  \cite{Ossola:2006us} the problem of reconstructing this rational
piece has been solved by looking at the implicit mass
dependence in the coefficients $d,c,b,a$ of the one-loop functions.
Such a method is adequate up to 4-point functions; for higher-point functions
the dependence becomes too complicated to be used in practice.
In addition, it requires the solution of further systems of linear
equations, slowing down the whole computation.
 For those reasons, we suggest here a different method.
One starts by rewriting any denominator appearing
in \eqn{eq:1} as follows
\bqa
\frac{1}{\db{i}} = \frac{\zb{i}}{\d{i}}\,,~~~~{\rm with}~~~
\zb{i}\equiv \left(1- \frac{\tld{q}^2}{\db{i}} \right)\,.
\eqa
This results in
\bqa
\label{eq:3}
A(\bar q)= \frac{N(q)}{\d{0}\d{1}\cdots \d{m-1}}\,
\zb{0} \zb{1} \cdots \zb{m-1}\,.
\eqa
Then, by inserting \eqn{eq:2} in \eqn{eq:3}, one obtains
\bqa
\label{eq:4}
A(\bar q) &=&
\sum_{i_0 < i_1 < i_2 < i_3}^{m-1}
\frac{
          d( i_0 i_1 i_2 i_3 ) +
     \tld{d}(q;i_0 i_1 i_2 i_3)
}{\db{i_0} \db{i_1} \db{i_2} \db{i_3}}
\prod_{i \ne i_0, i_1, i_2, i_3}^{m-1} \zb{i} \nl
     &+&
\sum_{i_0 < i_1 < i_2 }^{m-1}
\frac{
          c( i_0 i_1 i_2) +
     \tld{c}(q;i_0 i_1 i_2)
}{\db{i_0} \db{i_1} \db{i_2}}
\prod_{i \ne i_0, i_1, i_2}^{m-1} \zb{i} \nl
     &+&
\sum_{i_0 < i_1 }^{m-1}
\frac{
          b(i_0 i_1) +
     \tld{b}(q;i_0 i_1)
}{\db{i_0} \db{i_1}}
\prod_{i \ne i_0, i_1}^{m-1} \zb{i} \nl
     &+&
\sum_{i_0}^{m-1}
\frac{
          a(i_0) +
     \tld{a}(q;i_0)
}{\db{i_0}}
\prod_{i \ne i_0}^{m-1} \zb{i} \nl
     &+& \tld{P}(q)
\prod_{i}^{m-1} \zb{i}\,.
\eqa
The rational part of the amplitude is then produced,
after integrating over $d^nq$, by the
$\tld{q}^2$ dependence coming from the various $\zb{i}$ in \eqn{eq:4}.
It is easy to see what happens, for any value of $m$, by recalling
the generic $q$ dependence of the spurious terms.
In the renormalizable gauge one has \cite{Ossola:2006us}
\bqa
\label{eq:5}
\tld{P}(q) &=& 0\,, \nl
\tld{a}(q;i_0) &=& \tld{a}^\mu(i_0;1)(q+p_{i_0})_\mu\,, \nl
\tld{b}(q;i_0 i_1) &=&
\tld{b}^\mu(i_0 i_1;1)(q+p_{i_0})_\mu +
\tld{b}^{\mu\nu}(i_0 i_1;2)(q+p_{i_0})_\mu(q+p_{i_0})_\nu \,,\nl
\tld{c}(q;i_0 i_1 i_2) &=&
\tld{c}^\mu(i_0 i_1 i_2;1)(q+p_{i_0})_\mu +
\tld{c}^{\mu\nu}(i_0 i_1 i_1;2)(q+p_{i_0})_\mu(q+p_{i_0})_\nu \,,\nl
 &&\!+~ \tld{c}^{\mu\nu\rho}(i_0 i_1 i_1;3)
(q+p_{i_0})_\mu(q+p_{i_0})_\nu(q+p_{i_0})_\rho \,,\nl
\tld{d}(q;i_0i_1i_2i_3) &=&
\tld{d}^\mu(i_0i_1i_2i_3;1)(q+p_{i_0})_\mu\,.
\eqa
\eqn{eq:5} simply states the fact that $\tld{a}(q;i_0)$ and
$\tld{d}(q;i_0 i_1 i_2 i_3)$
are at most linear in $(q+p_{i_0})$,
$\tld{b}(q;i_0 i_1)$ at most quadratic, and $\tld{c}(q;i_0 i_1 i_2)$
at most cubic.
The tensors denoted by $(\cdots; 1)$, $(\cdots; 2)$
and  $(\cdots; 3)$ stand for the respective coefficients.
We will also make use of the fact that, due to the explicit form
of the spurious terms \cite{Ossola:2006us}
\bqa
\label{eq:7}
\tld{c}^{\mu\nu}( i_0 i_1 i_2; 2 )     \,g_{\mu \nu} &=& 0\,, \nl
\tld{c}^{\mu\nu\rho}( i_0 i_1 i_2; 3 ) \,g_{\mu \nu}  &=&
\tld{c}^{\mu\nu\rho}( i_0 i_1 i_2; 3 ) \,g_{\mu \rho}=
\tld{c}^{\mu\nu\rho}( i_0 i_1 i_2; 3 ) \,g_{\nu \rho}=  0~~{\rm and} \nl
\tld{b}^{\mu\nu}( i_0 i_1; 2 )     \,g_{\mu \nu} &=& 0\,.
\eqa
The necessary integrals that arise, after a change of
variable $q \to q - p_{i_0}$, are of the form
\bqa
\label{eq:6}
I^{(n;2\ell)}_{s;\mu_1 \cdots \mu_r} &\equiv& \int d^nq\,\tld{q}^{2 \ell} 
\frac{q_{\mu_1} \cdots q_{\mu_r}}{\bar D(k_{0}) \cdots \bar D(k_{s})}
\,,~~~{\rm with} \nl
\bar D(k_i) &\equiv& ({\bar q} + k_i)^2-m_i^2\,,~~~
k_i \equiv p_i -p_0\,~~~(k_{0} = 0)\,,
\eqa
where we used a notation introduced in \cite{delAguila:2004nf}
and $r \le 3$. Such integrals (from now on called extra-integrals)
have dimensionality ${\cal D}= 2(1+\ell -s)+r$
and give a contribution ${\cal} O(1)$ only when ${\cal D} \ge 0$,
otherwise are of ${\cal O}(\epsilon)$. 
This counting remains valid also in the presence of infrared and
collinear divergences, as explained, for example, in
Appendix B of \cite{delAguila:2004nf} and in \cite{Binoth2}.

We also note that, since all $\zb{i}$ are a-dimensional,
the dimensionality ${\cal D}$ of the extra-integrals
generated through \eqn{eq:4} does not depend on $m$.
We list, in the following, all possible contributions,
collecting the computational details in \app{appa}.

\vspace{0.3cm}

\noindent {\bf Contributions proportional to $d( i_0 i_1 i_2 i_3 )$}

\vspace{0.3cm}

\noindent In this case $r= 0$. All extra-integrals are therefore scalars
with ${\cal D}= -4$ and do not contribute.

\vspace{0.3cm}

\noindent {\bf Contributions proportional to
$\tld{d}^\mu ( i_0 i_1 i_2 i_3; 1 )$}

\vspace{0.3cm}

\noindent In this case $r= 1$. All extra-integrals
are therefore rank one tensors with ${\cal D}= -3$ and do not contribute.

\vspace{0.3cm}

\noindent {\bf Contributions proportional to
$c( i_0 i_1 i_2)$}

\vspace{0.3cm}

\noindent In this case $r= 0$ with ${\cal D}= -2$ and no contribution
${\cal O}(1)$ is developed.

\vspace{0.3cm}

\noindent {\bf Contributions proportional to
$\tld{c}^\mu( i_0 i_1 i_2; 1 )$}

\vspace{0.3cm}

\noindent Here $r= 1$ and ${\cal D}= -1$. Therefore, once again,
there is no contribution.

\vspace{0.3cm}

\noindent {\bf Contributions proportional
to $\tld{c}^{\mu\nu}( i_0 i_1 i_2; 2 )$}

\vspace{0.3cm}

\noindent Now $r= 2$ with ${\cal D}= 0$ and a finite contribution is
in principle expected, generated by extra-integrals of the type
\bqa
I^{(n;2(s-2))}_{s;\mu \nu}\,.
\eqa
Nevertheless, such contribution is proportional to $g_{\mu \nu}$
\cite{delAguila:2004nf}. Therefore, due to \eqn{eq:7},
it vanishes.

\vspace{0.3cm}

\noindent {\bf Contributions proportional
to $\tld{c}^{\mu\nu\rho}( i_0 i_1 i_2; 3 )$}

\vspace{0.3cm}

\noindent Now $r= 3$ and  ${\cal D}= 1$. The contributing
extra-integrals are of the type
\bqa
I^{(n;2(s-2))}_{s;\mu \nu \rho}\,,
\eqa
and one easily proves that the
contributions ${\cal O}(1)$ are always proportional
to $g_{\mu \nu}$ or $g_{\mu \rho}$ or $g_{\nu \rho}$.
Therefore, thanks again to \eqn{eq:7}, they vanish.

\vspace{0.3cm}

\noindent {\bf Contributions proportional
to $b( i_0 i_1)$}

\vspace{0.3cm}

\noindent Those are the first non vanishing contributions.
The relevant extra-integrals have
$r= 0$ and  ${\cal D}= 0$
\bqa
I^{(n;2(s-1))}_{s}\,, \nonumber
\eqa
with $ 2 < s \le m-1$.
They have been computed, for generic values of $s$,
 in \cite{delAguila:2004nf} (see also \app{appa})
\bqa
\label{eq:ints0}
I^{(n;2(s-1))}_{s}= -i \pi^2 \frac{1}{s(s-1)} + {\cal O}(\epsilon)\,.
\eqa

\vspace{0.3cm}

\noindent {\bf Contributions proportional
to $\tld{b}^\mu( i_0 i_1;1)$}

\vspace{0.3cm}

\noindent In this case the relevant extra-integrals are 4-vectors with
${\cal D}= 1$
\bqa
I^{(n;2(s-1))}_{s;\mu}~~~{\rm with}~~2 < s \le m-1\,.
\nonumber
\eqa
A computation for generic values of $s$ gives
\bqa
\label{eq:intv1}
I^{(n;2(s-1))}_{s;\mu}= i \pi^2 \frac{1}{(s+1)s(s-1)}
\sum_{j=1}^s (k_j)_\mu + {\cal O}(\epsilon)\,.
\eqa

\vspace{0.3cm}

\noindent {\bf Contributions proportional
to $\tld{b}^{\mu\nu}( i_0 i_1;2)$}

\vspace{0.3cm}

\noindent The relevant extra-integrals are now rank two tensor with
${\cal D}= 2$
\bqa
I^{(n;2(s-1)}_{s;\mu\nu}~~~{\rm with}~~2 < s \le m-1\,.
\nonumber
\eqa
They read
\bqa
\label{eq:intt2}
I^{(n;2(s-1))}_{s;\mu\nu} &=& -2 i \pi^2 \frac{1}{(s+2)(s+1)s(s-1)}
\left\{
\sum_{j=1}^s (k_j)_\mu (k_j)_\nu
+\frac{1}{2} \sum_{j=1}^s  \sum_{i \ne j}^s  (k_j)_\mu (k_i)_\nu
\right\}\, \nl
&+& {\cal O}(g_{\mu\nu})
+ {\cal O}(\epsilon)\,.
\eqa
The $g_{\mu\nu}$ part is never needed because
$\tld{b}^{\mu\nu}( i_0 i_1; 2 ) \,g_{\mu \nu} = 0$, according to \eqn{eq:7}.

\vspace{0.3cm}

\noindent {\bf Contributions proportional
to $a( i_0)$}

\vspace{0.3cm}

\noindent
They involve scalar extra-integrals with ${\cal D}= 2$
\bqa
I^{(n;2s)}_{s}\,,
~~~{\rm with}~~1 < s \le m-1\,.
\nonumber
\eqa
One computes
\bqa
\label{eq:ints2}
I^{(n;2s)}_{s} &=& - 2 i \pi^2 \frac{1}{(s+2)(s+1)s}
\left\{
\sum_{j=1}^s k_j^2
+\frac{1}{2} \sum_{j=1}^s  \sum_{i \ne j}^s  (k_j \cdot k_i)
+ \frac{s+2}{2} \sum_{j=0}^s (m_j^2-k_j^2)
\right\}\, \nl
 &+& {\cal O}(\epsilon)\,.
\eqa

\noindent {\bf Contributions proportional
to $\tld{a}^\mu ( i_0;1)$}

\vspace{0.3cm}

\noindent
This last category involves extra-integrals with
$r = 1$ and  ${\cal D}= 3$
\bqa
I^{(n;2s)}_{s;\mu}\,,
~~~{\rm with}~~1 < s \le m-1\,.
\nonumber
\eqa
One obtains
\bqa
\label{eq:intv3}
I^{(n;2s)}_{s;\mu} &=& i \pi^2 \frac{1}{(s+3)(s+2)(s+1)s}
\left\{
6 \sum_{j=1}^s k_j^2 (k_j)_\mu
+ 2 \sum_{j=1}^s  \sum_{i \ne j}^s
\left[
k_j^2 (k_i)_\mu +2 (k_j \cdot k_i) (k_j)_\mu
\right] \right. \nl
&+& \left. \sum_{j=1}^s  \sum_{i \ne j}^s  \sum_{\ell \ne i}^s
(k_j \cdot k_i) (k_\ell)_\mu
+(s+3)
\left[
2 \sum_{j=0}^s (m_j^2 - k_j^2) (k_j)_\mu  \right. \right. \nl
&+& \left. \left.
\sum_{j=0}^s  \sum_{i \ne j}^s (m_j^2 - k_j^2) (k_i)_\mu
\right]
\right\} + {\cal O}(\epsilon)\,.
\eqa

\vspace{0.3cm}

\noindent To conclude, the set of the five formulas in
\eqnss{eq:ints0}{eq:intv3}
allows one to compute the rational part of any one-loop $m$-point
(sub-)amplitude, once all the coefficients of
\eqn{eq:2} have been reconstructed.

\section{Dealing with numerical instabilities\label{numinacc}}
In this section we show how to handle, in the framework
of the method illustrated in the previous section,
the simplest numerical instability appearing in any one-loop calculation,
namely that one related to the tensor reduction of 2-point amplitudes in the
limit of vanishing Gram-determinant
\footnote{In this case the Gram-determinant
is simply the square of the difference between
the momenta of the two denominators.}.
This situation is simple enough to allow an easy description, but the
outlined strategy is general and not restricted to the 2-point case.

We start from the {\it integrand} of a generic 2-point amplitude
written in the form
\bqa
\label{eq:amp2}
A(\bar q)= \frac{N(q)}{\db{0}\db{1}}\,,
\eqa
in which we suppose $N(q)$ at most quadratic in $q$.
Our purpose is dealing with the situation in which
$k_1^2 \equiv (p_1- p_0)^2= 0 $ exactly (that always occur in processes with
massless external particles), as well as to set up an algorithm to write down
approximations around this case with arbitrary precision.

According to \eqn{eq:2}, we can write an expansion for $N(q)$ as follows:
\bqa
\label{eq:2a}
N(q) &=&
[
          b(01) +
     \tld{b}(q;01)
]
     +
[
          a(0) +
     \tld{a}(q;0)
] \d{1}
+
[
          a(1) +
     \tld{a}(q;1)
] \d{0}\,.
\eqa
If the Gram-determinant of the 2-point function is small, the reduction method
introduced in \cite{Ossola:2006us} cannot be applied, because
the solution for which $\d{0}= \d{1}= 0$, needed
to determine the coefficients $b$ and $\tld{b}$,
becomes singular\footnote{Such a solution goes like $1/k_1^2$.}, in
the limit of $k_1^2 \to 0$,
when adding the requirement
\bqa
\label{eq:spur0}
\int\,d^n q\, \tld{b}(q;01)  = 0\,.
\eqa
Then, we must consider two separate cases:
\bqa
\label{eq:cases}
k_1^2 &\to& 0\,,~~{\rm but}~~   k_1^\mu \ne 0\,, \nl
k_1^2 &\to& 0\,,~~{\rm because}~~k_1^\mu = 0\,.
\eqa
The former situation may occur because
of the Minkowskian metric, while the latter
takes place at the edges of the phase-space,
where some momenta become collinear.
In the first case one can still find a solution
for which $\d{0}= \d{1}= 0$ by relaxing the further requirement
of \eqn{eq:spur0}. Such a solution is given in \app{appb}
and goes like $1/(k_1.v)$, where $v$ is an arbitrary massless
4-vector, therefore is never singular in the first case of \eqn{eq:cases}.
The price to pay is that new non zero integrals appear of the type
\footnote{Since $v^2= 0$ they still fulfill the third one of
Eqs.~(\ref{eq:7}), therefore, even in this case, 
terms ${\cal O} (g_{\mu\nu})$ can be neglected in
\eqn{eq:intt2}.}
\bqa
\label{eq:newint}
\int\,d^n q\frac{[(q+p_0)\cdot v]^j}{\db{0} \db{1}}
~~~~{\rm with}~~~j= 1,2~~{\rm and}~~v^2= 0\,.
\eqa
What has been achieved with this new basis is then moving part
of the 1-point functions to the 2-point sector, in such a way that
combinations well behaved in the limit $k_1^2 \to 0$ appear.
The fact that solutions exist to the condition
$\d{0}= \d{1}= 0$, still allows one to find the
coefficients of such integrals (together with all the others).
This solves the first part of the problem, namely reconstructing
$N(q)$ without knowing explicitly its analytic structure, but
one is left with the problem of computing the new
2-point integrals. In the following, we present our
method to determine them at any desired order in $k_1^2$.
Let us first consider the case $j=1$ in \eqn{eq:newint}.
The contribution ${\cal O}(1)$ can be easily obtained from
the observation that\footnote{From now on, we shift the
integration variable: $\bar q \to \bar q-p_0$.
The definition of the new resulting denominators is given in \eqn{eq:6}.
}
\bqa
\label{eq:firstint}
\int\,d^n q\, \frac{(q\cdot v)(q \cdot k_1)^2}{\bar D(k_0) \bar D(k_1)}
= {\cal O} (k_1^2)\,,
\eqa
as it is evident by performing a tensor decomposition.
On the other hand, by reconstructing denominators, one obtains
\bqa
\label{eq:expan2}
(q \cdot k_1)^2 = \left(\frac{f}{2}\right)^2
+\frac{\bar D(k_1)-\bar D(k_0)}{2}\left[(q \cdot k_1)+\frac{f}{2}
\right]\,,
\eqa
with
\bqa
\label{eq:f}
f = m_1^2-k_1^2-m_0^2\,.
\eqa
\eqn{eq:expan2}, inserted in \eqn{eq:firstint}
gives the desired expansion in terms of loop functions with less points but
higher rank, in agreement with well know results
\cite{Devaraj:1997es,Denner:2006fy}
\bqa
\int\,d^n q\frac{(q\cdot v)}{\bar D(k_0) \bar D(k_1)}=
\frac{1}{f}
\int \,d^n q\,  (q\cdot v)
\left( \frac{1}{\bar D(k_1)}-\frac{1}{\bar D(k_0)} \right)
\left(1+ \frac{2(q\cdot k_1)}{f}\right) +
{\cal O} (k_1^2)\,.
\eqa
Expansions at arbitrary orders in $k_1^2$ can be obtained in an analogous way
from the two following equations:
\bqa
&&(q \cdot k_1)^{p} = \left(\frac{f}{2}\right)^{p}
+\frac{\bar D(k_1)-\bar D(k_0)}{2}\sum_{i+j= p-1}
\left[(q \cdot k_1)^i \left(\frac{f}{2}\right)^j  \right]\,, \nl
&&\int\,d^n q\frac{(q\cdot v)(q \cdot k_1)^{2p}}
{\bar D(k_0) \bar D(k_1)} =  {\cal O} (k_1^{2p})\,.
\eqa
To deal with the case
$j= 2$ in \eqn{eq:newint} one starts instead from
the equation
\bqa
\int\,d^n q\frac{(q\cdot v)^2(q \cdot k_1)^{2p+1}}{\bar D(k_0) \bar D(k_1)}
={\cal O} (k_1^{2p})\,.
\eqa
This procedure breaks down when the quantity $f$ vanishes.
In this case a double expansion
in $k_1^2$ and $f$  can still be found in terms of derivatives
of one-loop scalar functions.
We illustrate the procedure for the case $j= 1$ of \eqn{eq:newint}.
Our starting point is now the equation
\bqa
\label{eq:feq0}
\bar D(k_0) = \bar D(k_1) -2 (q \cdot k_1) + f\,.
\eqa
By multiplying and dividing by $\bar D(k_0)$ one obtains
\bqa
\int\,d^n q\frac{(q\cdot v)}{\bar D(k_0) \bar D(k_1)} &=&
\int\,d^n q\frac{(q\cdot v)}{\bar D(k_0)^2 \bar D(k_1)}
\left[
\bar D(k_1) -2 (q \cdot k_1) + f
\right] \nl
&=&
\int\,d^n q\frac{(q\cdot v)}{\bar D(k_0)^2}
-2 \int\,d^n q\frac{(q\cdot v)(q\cdot k_1)}{\bar D(k_0)^2 \bar D(k_1)}
+ {\cal O}(f)\,.
\eqa
Applying once more \eqn{eq:feq0} to the last integral gives
\bqa
 \int\,d^n q\frac{(q\cdot v)(q\cdot k_1)}{\bar D(k_0)^2 \bar D(k_1)}
&=& \int\,d^n q\frac{(q\cdot v)(q\cdot k_1)}{\bar D(k_0)^3 \bar D(k_1)}
\left[
\bar D(k_1) -2 (q \cdot k_1) + f
\right] \nl
&=&
 \int\,d^n q\frac{(q\cdot v)(q\cdot k_1)}{\bar D(k_0)^3}
-2 \int\,d^n q\frac{(q\cdot v)(q\cdot k_1)^2}{\bar D(k_0)^3 \bar D(k_1)}
+ {\cal O}(f)\,.
\eqa
Since the last integral in the previous equation is ${\cal O}(k_1^2)$,
the final result reads
\bqa
 \int \,d^n q \,\frac{(q\cdot v)}{\bar D(k_0) \bar D(k_1)} &=&
 \int \,d^n q \, \frac{(q\cdot v)}{\bar D(k_0)^2}
-2\int \,d^n q \,\frac{(q\cdot v)(q \cdot k_1)}{\bar D(k_0)^3}
+{\cal O}(k_1^2)+{\cal O}(f) \,. \nl
\eqa
In a similar fashion, expansions at any order can be obtained.

We now turn to the second case of \eqn{eq:cases}, namely $k_1^\mu \to 0$.
In this case {\em no solution} can be found to the double cut equation
\bqa
D(k_0)= D(k_1)= 0\,.
\eqa
The reason is that now $D(k_1)$ and $D(k_0)$ are no longer independent:
\bqa
D(k_0)= D(k_1) + f +{\cal O}(k_1)\,,
\eqa
and clearly no $q$ exists such that the two denominators
can be simultaneously zero.
Notice that this also implies that one cannot fit
separately the coefficients of the 2-point and 1-point
functions in \eqn{eq:2a}.
This results is a singularity $1/(k_1\cdot v)$ in the system
given of \app{appb} and we should change our strategy.
We than go back to \eqn{eq:amp2} and split the amplitude from the beginning
by multiplying it by
\bqa
\label{eq:one}
1 \equiv \frac{\bar D(k_0)-\bar D(k_1)}{f}
+\frac{2 (q \cdot k_1)}{f}\,,
\eqa
resulting to
\bqa
\label{eq:asplit}
A(\bar q)= A^{(1)}(\bar q)+A^{(2)}(\bar q) + {\cal O}(k_1)\,,
\eqa
with
\bqa
A^{(1)}(\bar q)=     \frac{1}{f}\frac{N(q)}{\bar D(k_1)}\,,~~~
A^{(2)}(\bar q)=-\frac{1}{f}\frac{N(q)}{\bar D(k_0)}\,.
\eqa
Now the two amplitudes $A^{(1,2)}$ can be reconstructed separately, without
any problem of vanishing Gram-determinant.
Notice also that corrections at orders higher than ${\cal O}(k_1)$
are perfectly calculable by inserting again
\eqn{eq:one} in the term ${\cal O}(k_1)$ of \eqn{eq:asplit}.

Once again, when $f \to 0$, double expansions
in $k_1$ and $f$ can be obtained involving derivatives of scalar
loop functions by using \eqn{eq:feq0}.
For example, at the zeroth order in $k_1$ and at the first one
in $f$, one gets
\bqa
A(q) &=& \frac{N(q)}{\bar D(k_0) \bar D(k_1)}=
\frac{N(q)}{\bar D(k_0)^2 \bar D(k_1)}
\left[\bar D(k_1)-2 (q \cdot k_1)+ f\  \right] \nl
 &=&
\frac{N(q)}{\bar D(k_0)^2}
+ f \frac{N(q)}{\bar D(k_0)^3 \bar D(k_1)}
\left[\bar D(k_1)-2 (q \cdot k_1)+ f\  \right]
+ {\cal O}(k_1) \nl
&=& \frac{N(q)}{\bar D(k_0)^2}
+ f\frac{N(q)}{\bar D(k_0)^3}
+ {\cal O}(k_1)
+ {\cal O}(f^2)\,.
\eqa
This last case exhausts all possibilities.

The same techniques can be applied for higher-point functions.
For example, in the case of a 3-point function, instead of
$k_1$, one introduces the 4-vector
\bqa
s^\mu = {\rm det}
\left|
\begin{tabular}{cc}
 $k_1^\mu$         & $k_2^\mu$ \\
 $(k_2 \cdot k_1)$ & $(k_2 \cdot k_2)$
\end{tabular}
\right|\,,
\eqa
with the properties
\bqa
s \cdot k_2 = 0\,,~~
s^2 \propto \Delta(k_1,k_2)\,,~~
(k_1 \cdot s) \propto \Delta(k_1,k_2)\,,
\eqa
where $\Delta(k_1,k_2)$ is the Gram-determinant of the two momenta
$k_1$ and $k_2$.
Then, instead of \eqn{eq:firstint} one has, for example,
\bqa
\int\,d^n q\, \frac{(q\cdot v)(q \cdot s)^2}{\bar D(k_0)\bar D(k_1)\bar D(k_2)}
= {\cal O} (\Delta(k_1,k_2))\,.
\eqa
As before, $\Delta(k_1,k_2)$ can vanish either because $s^2 = 0$ or
$s^\mu =0$ and the two cases should be treated separately.

\section{Results and comparisons \label{numerics}}
We started by checking our implementation of the rational terms. For
4-point functions up to rank four, we reproduced the results
obtained with the alternative technique illustrated in
\cite{Ossola:2006us}. Furthermore, we reproduced the rational part
of the full $2 \gamma \to 2 \gamma$ amplitude given in
\cite{Gounaris:1999gh}. We also checked with an independent
calculation \cite{vanHameren:2005ed} the rational terms coming from
all of the 6-point tensors up to rank six. Finally, we computed the
rational piece of the whole $2 \gamma \to 4 \gamma$ amplitude by
summing up all 120 contributing Feynman diagrams and finding zero,
as it should be \cite{Binoth2}.

As a first test on full amplitudes, we checked our method by reproducing
the contribution of a fermion loop to the $2 \gamma \to 2 \gamma$
process. This result is presented in Eqs.~(A.18)-(A.20)
of Ref.~\cite{Gounaris:1999gh},
for all possible helicity configurations.
We are in perfect agreement with the analytic result, in both massless
and massive cases.

The next step was the computation of the $2 \gamma \to 4 \gamma$
amplitude with zero internal mass\footnote{We thank Andre van
Hameren for providing us with his program to compute massless
one-loop scalar integrals.}, finding the results given in
\fig{plot1} and \fig{plot2}. It should be mentioned that our results
are obtained algebraically, so there is no integration error
involved.
%
\FIGURE{\begin{rotate}{90}\hspace*{30mm}{\small
$s |{\cal M}|/\alpha^3 $}\end{rotate}
\epsfig{file=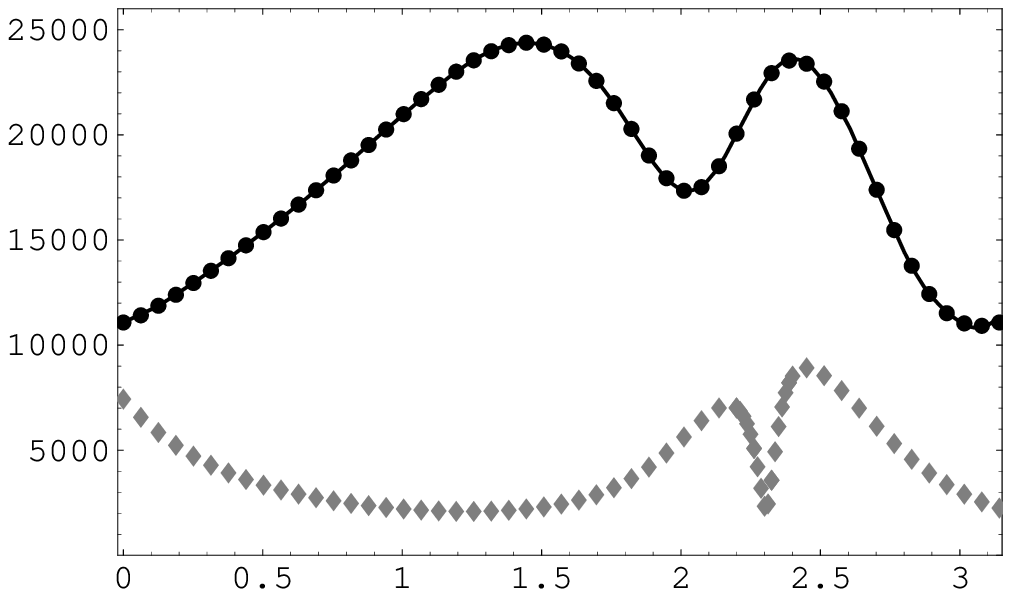,height=7cm,width=9cm}
$\theta$
\caption{\label{plot1} Comparison with Fig.~5 of Ref.~\cite{nagy}.
Helicity configurations \mbox{$[++----]$} and  \mbox{$[+--++-]$}
for the momenta of \eqn{mom1}, represented by black dots
and gray diamonds respectively, and comparison
with the analytic result of Ref.~\cite{mahlon} (continuous line).}}
%
\FIGURE{\begin{rotate}{90}\hspace*{30mm}{\small
$s |{\cal M}|/\alpha^3 $}\end{rotate}
\epsfig{file=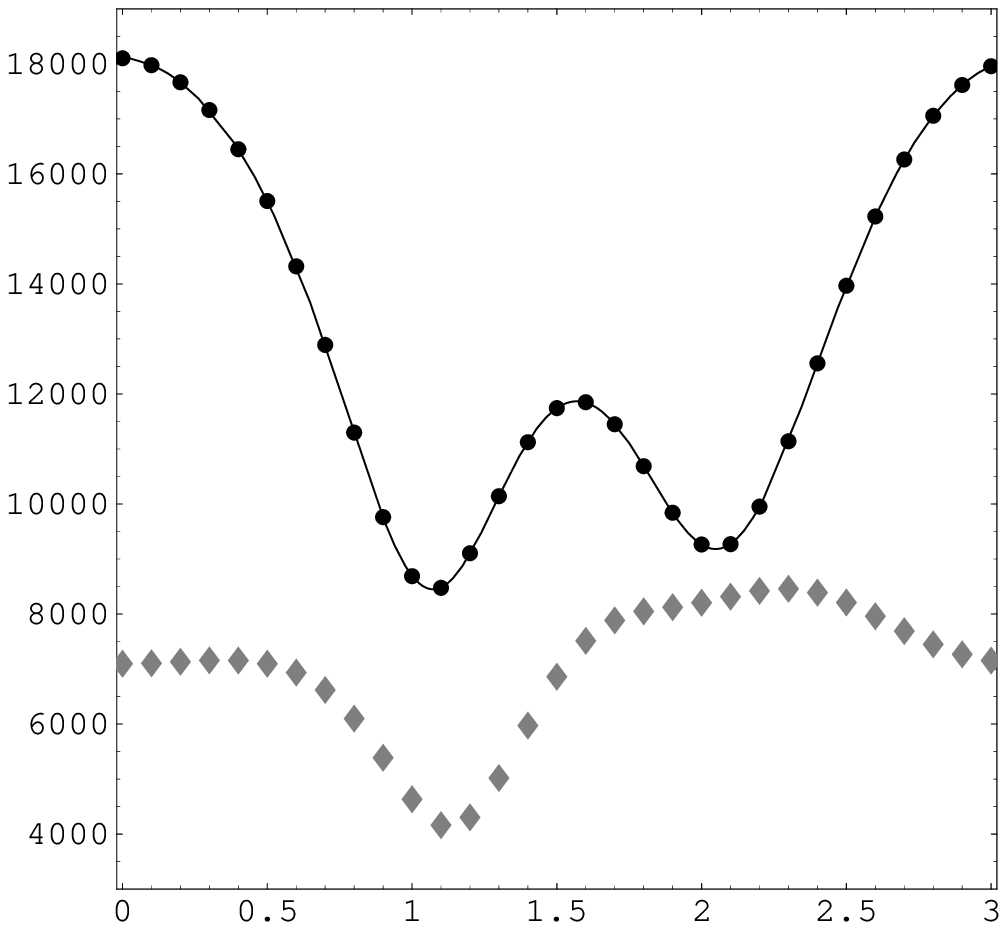,height=7cm,width=9cm}
$\theta$
\caption{\label{plot2}
Helicity configurations \mbox{$[++----]$} and  \mbox{$[++--+-]$}
for the momenta of \eqn{mom2}, represented by black dots and gray diamonds respectively, and comparison
with the analytic result of Ref.~\cite{mahlon} (continuous line).}}
In \fig{plot1}, we reproduce the
results presented by Nagy and Soper \cite{nagy} and very recently
also by Binoth et al.\cite{sixph}.
We employ the same values of the external momenta as in Fig. 5
of Ref.~\cite{nagy}, namely the following selection
of final state three-momenta $\{\vec p_3, \vec p_4, \vec p_5, \vec p_6\}$:
\bqa
\label{mom1}
\vec p_3 &=&  (33.5,15.9,25.0  ) \,, \nl
\vec p_4 &=&  (-12.5,15.3,0.3)   \,, \nl
\vec p_5 &=& (-10.0,-18.0,-3.3)  \,, \nl
\vec p_6 &=& (-11.0,-13.2,-22.0) \,.
\eqa
After choosing the incoming photons such that they have momenta
$\vec p_1$ and  $\vec p_2$ along the $z$-axis, we present in the plot
the amplitude obtained by rotating the final states of angle $\theta$
about the $y$-axis.
This is done for both helicity configurations \mbox{$[++----]$} and
\mbox{$[+--++-]$}.
In the same plot also appears the analytic results
for the configuration \mbox{$[++----]$} obtained by Mahlon \cite{mahlon}.
In \fig{plot2}, we use a different
set of external momenta. Starting from the following
choice of $\{\vec p_3, \vec p_4, \vec p_5, \vec p_6\}$:
\bqa
\label{mom2}
\vec p_3 &=&  (-10.0, -10.0, -10.0) \,, \nl
\vec p_4 &=&  (12.0, -15.0, -2.0)   \,, \nl
\vec p_5 &=& (10.0, 18.0, 3.0)      \,, \nl
\vec p_6 &=& (-12.0, 7.0, 9.0)      \,
\eqa
we proceed as in the previous case. The results for the amplitudes
are plotted in \fig{plot2} for the helicity configurations
\mbox{$[++----]$} and  \mbox{$[++--+-]$}.
It is known that the six-photons amplitude vanish
for the helicity configurations \mbox{$[++++++]$} and \mbox{$[+++++-]$},
we checked this result for both choices of the external momenta.
Finally, using the external momenta of \eqn{mom1}, we computed
the amplitude introducing a non-zero mass $m_f$  for the fermions in the loop
\footnote{We used here the scalar one-loop functions provided by FF\cite{FF}.}.
The results are plotted in \fig{plot3}, for the three cases
$m_f = 0.5$ GeV, $m_f= 4.5$ GeV and $m_f= 12$ GeV.

The code we prepared for producing the results presented in this section is
written in FORTRAN 90. Even if we did not spend too much effort 
in optimizations, it can compute about 3 phase-space points per second,
when working in double precision. All figures in this section
are actually produced by using double precision, but, to perform a realistic
integration, we still need quadruple precision, that slows down the speed by
about a factor 60. We are working in implementing the expansions 
presented in the previous section with the aim
of being able to perform a stable integration over the full phase space, 
that is a ``proof of concept'' for any method.

\FIGURE{\begin{rotate}{90}\hspace*{30mm}{\small
$s |{\cal M}|/\alpha^3 $}\end{rotate}
\epsfig{file=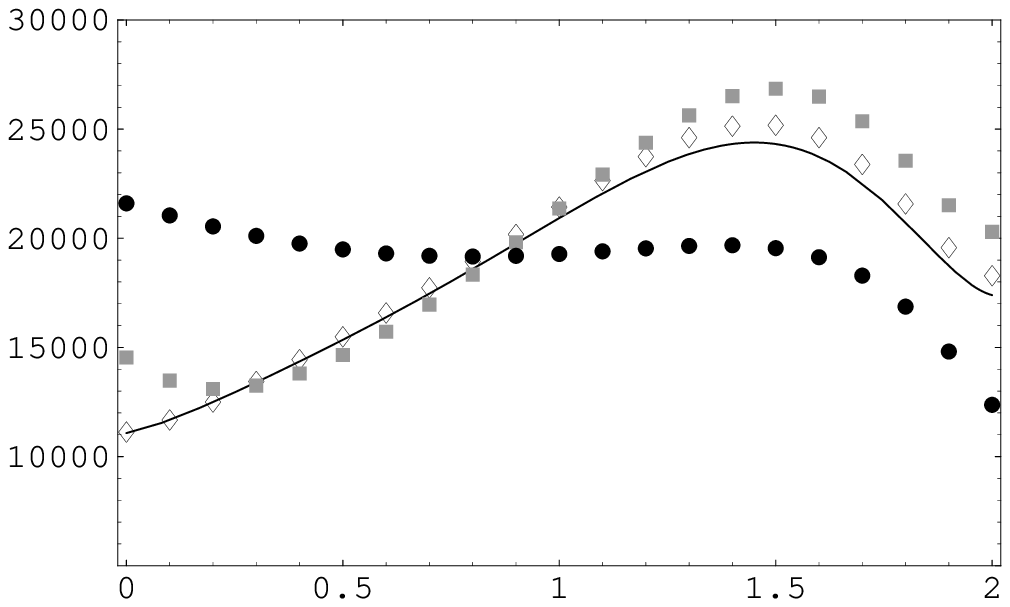,height=7cm,width=9cm}
$\theta$
\caption{\label{plot3}
Helicity configuration \mbox{$[++----]$}
for the momenta of \eqn{mom1} for different values of the fermion mass
in the loop: $m_f = 0.5$ GeV (diamond), $m_f= 4.5$ GeV (gray box)  
and $m_f= 12$ GeV (black dots). The continuous line is the result 
for the massless case.}}

\section{Conclusions}
Computing the massless QED amplitude for the reaction $2 \gamma \to
4 \gamma $, although still unobserved experimentally, is a very good
exercise for checking new methods to calculate one-loop virtual
corrections.
 Such a process posses all complications typical of any
multi-leg final state, for example a non trivial tensorial structure,
but also keeps, at the same time, enough simplicity such that
compact analytical formulas can still be used as a benchmark.
However, it is oversimplified in two respects.
Firstly,  the amplitude it is completely massless.
Secondly, the amplitude is cut constructible,
namely does not contain any rational part.

In the most general case of one-loop calculations, the presence of
both internal and external masses prevents from obtaining
compact analytical expressions. Then one has to rely on other
computational techniques. For example, it is known that
cut-constructible amplitudes can be obtained through recursion
relations. But, even then, the presence of rational parts usually
requires a separate work.

 For such reasons, it would be highly advisable to have
a method not restricted to massless theories, in which moreover both
cut-constructible and rational parts can be treated {\it at the same
time}. Such a method has been introduced recently in
Ref.~\cite{Ossola:2006us} and, in this paper, we applied it to the
computation of the six-photon amplitude in QED, giving also results
for the case with massive fermions in the loop. We also showed in
detail how the rational part of any $m$-point one-loop amplitude is
intimately connected with the form of the {\it integrand} of the
amplitude. Once this {\it integrand} is numerically computable,
cut-constructible and rational terms are easily obtained, at the
same time, by solving the same system of linear equations. This is a
peculiar property of our method, that we tested in the actual
computation of the six-photon amplitude. In practice, we did not use
the additional information on its cut-constructibility and verified
only {\it a-posteriori} that the intermediate rational parts, coming
from all pieces separately, drop out in the final sum.

Finally, we presented all relevant formulas needed to infer the
rational parts from the {\it integrand} of any $m$-point loop
functions, in the renormalizable gauges.

In addition, we presented, by analyzing in detail the 2-point
case, an idea to cure the numerical instabilities occurring
at exceptional phase-space points, outlining a possible way
to build up expansions around the zeroes of the Gram-determinants.

Having been able to apply our method to
the computation of the massive six-photon amplitude,
we are confident that our method can be successfully
used for a systematic and efficient
computation of one-loop amplitudes relevant
at LHC and ILC.

\section*{Acknowledgments}
We thank Andre van Hameren for numerical comparisons and Zoltan Nagy and
Pierpaolo Mastrolia for interesting discussions.
G.O. acknowledges the financial support of the ToK
Program ``ALGOTOOLS'' (MTKD-CT-2004-014319).
C.G.P's and  R.P.'s research was partially supported by the RTN European
Programme MRTN-CT-2006-035505 (HEPTOOLS, Tools and Precision Calculations for
Physics Discoveries at Colliders).
The research of R.P. was
also supported in part by MIUR under contract 2006020509\_004.

\section*{Appendices}
\appendix
\section{Computing the extra-integrals \label{appa}}
In this appendix, we compute
the extra-integrals listed in Section \ref{ratpart}.
Since a contribution ${\cal O}(1)$
can only develop for non-negative dimensionality ${\cal D}$,
the integrand in the Feynman parameter integral is always polynomial.
First we decompose the integration as follows
\bqa
\int d^n \bar q= \int d^4q\, d^\epsilon \mu~~~~~(\tilde{q}^2= -\mu^2)\,,
\eqa
then, after using Feynman parametrization and performing first
the integral over $d^\epsilon \mu$ and then that one over $d^4q$,
one derives, for the extra-integrals of \eqnss{eq:ints0}{eq:intv3}
\bqa
\label{eq:intextr}
I^{(n;2(s-1))}_{s}  &=&  -i \pi^2  \Gamma(s-1) \int [d \alpha]_s
+ {\cal O(\epsilon)}\,, \nl
I^{(n;2(s-1))}_{s;\mu} &=&
i \pi^2 \Gamma(s-1)
\int [d \alpha]_s\, (P_s)_\mu + {\cal O(\epsilon)}\,,\nl
I^{(n;2(s-1))}_{s;\mu\nu} &=&  -i \pi^2 \Gamma(s-1)
\int [d \alpha]_s\, (P_s)_\mu (P_s)_\nu + {\cal O}(g_{\mu\nu})
+ {\cal O(\epsilon)}\,, \nl
I^{(n;2s)}_{s} &=&     -i \pi^2 \Gamma(s)
\int [d \alpha]_s\, {\cal X}_s + {\cal O(\epsilon)}\,, \nl
I^{(n;2s)}_{s;\mu} &=&  i \pi^2 \Gamma(s)
\int [d \alpha]_s\, {\cal X}_s (P_s)_\mu + {\cal O(\epsilon)}\,,
\eqa
where
\bqa
\label{eq:app4}
\int [d \alpha]_s &=& \int_0^{\infty} d\alpha_0 \cdots d\alpha_s~
\delta(1-\sum_{j= 0}^s \alpha_j)\,,~~~{\cal X}_s = P_s^2+M_s^2\,, \nl
P_s &=& \sum_{j= 0}^s \alpha_j k_j\,,
~~~M_s^2 = \sum_{j= 0}^s \alpha_j (m_j^2-k_j^2)\,,~~(k_0= 0)\,.
\eqa
In the following, we compute, as an illustrative example,
the first three integrals of \eqn{eq:intextr}.
The remaining two can be obtained analogously.
We start by changing the integration variables as follows:
\bqa
\alpha_1 &=& \rho_1 \rho_2 \cdots \rho_s \nl
\alpha_2 &=& \rho_1 \rho_2 \cdots \rho_{s-1} (1-\rho_s)   \nl
\alpha_3 &=& \rho_1 \rho_2 \cdots \rho_{s-2} (1-\rho_{s-1})\nl
.\nl
.\nl
\alpha_s &=& \rho_1 (1-\rho_2) \nl
\alpha_0 &=& (1-\rho_1)\,,
\eqa
so that
\bqa
\int [d \alpha]_s =
\int_0^1 d \rho_1 \int_0^1 d \rho_2 \cdots \int_0^1 d \rho_s
\,\rho_1^{(s-1)} \rho_2^{(s-2)}
\cdots \rho_{s-1}\,,
\eqa
from which one trivially obtains the first integral
\bqa
I^{(n;2(s-1))}_{s}  &=&  -i \pi^2  \frac{\Gamma(s-1)}{\Gamma(s+1)}
+ {\cal O(\epsilon)}\,.
\eqa
To compute the second integral an integration over
$(P_s)_\mu$ in needed. Since the integrand
is symmetric when interchanging all $k_i$,
we concentrate on the coefficient of, say, $k_1$.
Since
\bqa
&&\int_0^1 d \rho_1 \int_0^1 d \rho_2 \cdots \int_0^1 d \rho_s
\,\rho_1^{(s-1)} \rho_2^{(s-2)}
\cdots \rho_{s-1}\,\alpha_1 k_{1\mu}\nl
&=& k_{1\mu}\,
\int_0^1 d \rho_1 \int_0^1 d \rho_2 \cdots \int_0^1 d \rho_s
\,\rho_1^{(s)} \rho_2^{(s-1)}
\cdots \rho^2_{s-1} \rho_s \nl
&=& k_{1\mu} \frac{1}{\Gamma(s+2)}\,,
\eqa
the final result reads
\bqa
I^{(n;2(s-1))}_{s;\mu} &=&
i \pi^2 \frac{\Gamma(s-1)}{\Gamma(s+2)}\,
\sum_{j=1}^s (k_j)_\mu
 + {\cal O(\epsilon)}\,.
\eqa
To compute the third integral we need to integrate
over the product $(P_s)_\mu (P_s)_\nu$. Once
again, given the symmetry of the problem, we can focus
on the two contributions proportional to
$k_{1\mu} k_{1\nu}$ and $k_{1\mu} k_{2\nu}$.
The first one gives
\bqa
&&\int_0^1 d \rho_1 \int_0^1 d \rho_2 \cdots \int_0^1 d \rho_s
\,\rho_1^{(s-1)} \rho_2^{(s-2)}
\cdots \rho_{s-1}\,\alpha_1^2 k_{1\mu} k_{1\nu}\nl
&=& k_{1\mu} k_{1\nu}\,
\int_0^1 d \rho_1 \int_0^1 d \rho_2 \cdots \int_0^1 d \rho_s
\,\rho_1^{(s+1)} \rho_2^{(s)}
\cdots \rho^3_{s-1} \rho^2_s \nl
&=& k_{1\mu} k_{1\nu} \frac{2}{\Gamma(s+3)}\,,
\eqa
and the second reads
\bqa
&&\int_0^1 d \rho_1 \int_0^1 d \rho_2 \cdots \int_0^1 d \rho_s
\,\rho_1^{(s-1)} \rho_2^{(s-2)}
\cdots \rho_{s-1}\,\alpha_1 \alpha_2 k_{1\mu} k_{2\nu}\nl
&=& k_{1\mu} k_{2\nu}\,
\int_0^1 d \rho_1 \int_0^1 d \rho_2 \cdots \int_0^1 d \rho_s
\,\rho_1^{(s+1)} \rho_2^{(s)}
\cdots \rho^3_{s-1} \rho_s (1-\rho_s) \nl
&=& k_{1\mu} k_{2\nu} \frac{1}{\Gamma(s+3)}\,.
\eqa
Summing up all of the possibilities one obtains
\bqa
I^{(n;2(s-1))}_{s;\mu\nu} &=& -2 i \pi^2 \frac{\Gamma(s-1)}{\Gamma(s+3)} \,
\left\{
\sum_{j=1}^s (k_j)_\mu (k_j)_\nu
+\frac{1}{2} \sum_{j=1}^s  \sum_{i \ne j}^s  (k_j)_\mu (k_i)_\nu
\right\}\, \nl
&+& {\cal O}(g_{\mu\nu})
+ {\cal O}(\epsilon)\,.
\eqa
\section{The general basis for the 2-point functions \label{appb}}
In this appendix, we solve the problem of
reconstructing the coefficients of the 2-point part of the {\it integrand}
of any amplitude
\bqa
A(\bar q)= \frac{N(q)}{\db{0}\db{1}}\,,
\eqa
by assuming $N(q)$ at most quadratic in $q$ and
$k_1 \equiv (p_1 -p_0) \ne 0$.
In particular also the case of vanishing $k_1^2$ is included.
First, we introduce a massless arbitrary 4-vector $v$,
such that $ (v \cdot k_1) \ne 0$, that we use to rewrite
$k_1$ in terms of two massless 4-vectors (we also take  $\ell^2=0$)
\bqa
k_1= \ell + \alpha \,v\,,
\eqa
giving
\bqa
\gamma \equiv 2\,(k_1 \cdot v) = 2\,(\ell \cdot v)~~{\rm and}~~
\alpha= \frac{k_1^2}{\gamma}\,.
\eqa
Then, we introduce two additional independent massless  4-vectors
$\ell_{7,8}$ defined as
\bqa
\ell_7^\mu &=& <\ell| \gamma^\mu |v]\,,~~
\ell_8^\mu = <v| \gamma^\mu |\ell]\,,
\eqa
for which one finds
\bqa
(\ell_7 \cdot \ell_8) =  -2 \gamma\,,
\eqa
and we decompose
$q^\mu+ p_0^\mu$ in the basis of $k_1$, $v$, $\ell_7$ and $\ell_8$
\bqa
\label{eq:qexp}
q^\mu = -p_0^\mu+ y k_1^\mu+ y_v v^\mu + y_7 \ell_7^\mu + y_8 \ell_8^\mu\,,
\eqa
so that $N(q)$ takes the form
\bqa
\label{eq:sysgen}
N(q) &=& b +
  \hat b_0    [(q+p_0) \cdot v]
 +\hat b_{00} [(q+p_0) \cdot v]^2 +
 \tld{b}_{11}[(q+p_0)\cdot \ell_7]
+\tld{b}_{21}[(q+p_0)\cdot \ell_8] \nl
&+&\tld{b}_{12}[(q+p_0)\cdot \ell_7]^2
  +\tld{b}_{22}[(q+p_0)\cdot \ell_8]^2 \nl
&+& \tld{b}_{01}[(q+p_0)\cdot \ell_7][(q+p_0)\cdot v] \nl
&+& \tld{b}_{02}[(q+p_0)\cdot \ell_8][(q+p_0)\cdot v]
+ {\cal O} (D_0)
+ {\cal O} (D_1)\,.
\eqa
Notice that, because of the identity
\bqa
2 \,(q \cdot k_1) = D_1-D_0+(d_1-d_0)\,,~~
{\rm with}~~d_i= m_i^2 - p_i^2\,,
\eqa
any term proportional to $[(q+p_0) \cdot k_1]$
either contributes to the
constant term $b$ or it is included in the terms
${\cal O} (D_{0,1})$ we are neglecting
\footnote{We suppose to determine them  at  a later stage of the calculation.}.
The same happens for the combination
$[(q+p_0)\cdot \ell_7][(q+p_0)\cdot \ell_8]$.

To be able to determine all of the coefficients appearing in
\eqn{eq:sysgen}, disentangling completely the contributions
${\cal O} (D_{0,1})$, we look for a $q$ that fulfill the
requirement
\bqa
D_0 = D_1 = 0\,.
\eqa
For a $q$ written as in \eqn{eq:qexp} this implies the system
\bqa
&&y_7 y_8 = F_y \nl
&&y_v     = \frac{d_1-d_0 - 2 y k_1^2}{\gamma}\,,
\eqa
where
\bqa
F_y = - \frac{1}{4 \gamma}
\left(
m_0^2-y\,(d_1-d_0)+y^2 k_1^2
\right)\,.
\eqa
It is convenient to introduce two classes of solutions.
In the first class, that we call $q^{\pm}_{yk}$,
we take $y$ fixed and choose $y_7= \pm e^{i\pi/k}$.
In the second class, that we call $q^{\prime \pm}_{yk}$,
we take $y$ fixed but choose $y_8= \pm e^{i\pi/k}$.
The coefficients $b$, $\tld{b}_{11}$,  $\tld{b}_{21}$,
$\tld{b}_{12}$ and $\tld{b}_{22}$ can be obtained
by evaluating \eqn{eq:sysgen} at the values
\bqa
q^{\pm}_{01}\,,~q^{\pm}_{02}\,,~q^{\pm}_{03}\,,
\eqa
or
\bqa
q^{\prime \pm}_{01}\,,~q^{\prime \pm}_{02}\,,~q^{\prime \pm}_{03}\,.
\eqa
In the first case, the coefficients read
\bqa
\label{eq:firstcase}
b_0   = b\,,~~b_1   = -2 \gamma \tld{b}_{21}\,,~~
b_2   =  4 \gamma^2 \tld{b}_{22}\,,~~
b_{-1}= -2 \gamma F_0\tld{b}_{11}\,,~~
b_{-2}=  4 \gamma^2 F_0^2\tld{b}_{12} \,,
\eqa
with
\bqa
b_{\pm 1} &=& -\frac{1}{2} \left[
T^-(q_1)\pm i T^-(q_2)
\right]\,, \nl
 b_0       &=& \frac{T^+(q_1)+T^+(q_2)}{2}\,, \nl
 b_{\pm 2} &=& \left[
 \frac{T^+(q_1)-T^+(q_2)}{2}
 -e^{\pm 2 i \pi/3}(T^+(q_3)-b_0)
\right]
\frac{1}{1-e^{\mp 2 i \pi/3}}\,,
\eqa
and where
\bqa
T^\pm(q_k) \equiv \frac{N(q^+_{0k})
                    \pm N(q^-_{0k})}{2}\,.
\eqa
In the second case, one obtains instead
\bqa
\label{eq:secondcase}
b_0^\prime    = b\,,~~b_1^\prime    = -2 \gamma \tld{b}_{11}\,,~~
b_2^\prime    =  4 \gamma^2 \tld{b}_{12}\,,~~
b_{-1}^\prime = -2 \gamma F_0\tld{b}_{21}\,,~~
b_{-2}^\prime =  4 \gamma^2 F_0^2\tld{b}_{22} \,,
\eqa
with
\bqa
b^\prime_{\pm 1} &=& -\frac{1}{2} \left[
T^-(q^\prime_1)\pm i T^-(q^\prime_2)
\right]\,, \nl
 b^\prime_0       &=& \frac{T^+(q^\prime_1)+T^+(q^\prime_2)}{2}\,, \nl
 b^\prime_{\pm 2} &=& \left[
 \frac{T^+(q^\prime_1)-T^+(q^\prime_2)}{2}
 -e^{\pm 2 i \pi/3}(T^+(q^\prime_3)-b^\prime_0)
\right]
\frac{1}{1-e^{\mp 2 i \pi/3}}\,,
\eqa
and where
\bqa
T^\pm(q^\prime_k) \equiv \frac{N(q^{\prime+}_{0k})
                    \pm N(q^{\prime-}_{0k})}{2}\,.
\eqa
The reason why we have chosen two sets of solutions is that, in
some special kinematical configurations, $F_0$ can vanish. Therefore,
numerical stable solutions are obtained by taking
$\tld{b}_{21}$ and $\tld{b}_{22}$
from \eqn{eq:firstcase}, and $\tld{b}_{11}$ and $\tld{b}_{12}$
from \eqn{eq:secondcase}, while $b$ is well defined in both cases.

The coefficients $\hat b_0$ and $\hat b_{00}$
can be determined, in terms of additional solutions of the kind
$q^\pm_{\lambda 1}$ and $q^\pm_{\sigma 1}$,
by defining the combinations
\bqa
S(q) &\equiv& N(q)
  -b
-\tld{b}_{11}[(q+p_0)\cdot \ell_7]
-\tld{b}_{21}[(q+p_0)\cdot \ell_8] \nl
&-&\tld{b}_{12}[(q+p_0)\cdot \ell_7]^2
  -\tld{b}_{22}[(q+p_0)\cdot \ell_8]^2\,, \nl \nl
U(\lambda) &\equiv& \frac{S(q^+_{\lambda 1}) + S(q^-_{\lambda 1})}{2}\,,
\eqa
as the two solutions of the system
\bqa
\left(
\begin{tabular}{l}
$U (\lambda)$ \\
$U (\sigma)$
\end{tabular}
\right)
=
\left(
\begin{tabular}{ll}
$\frac{\lambda\gamma}{2}$   & $\frac{\lambda^2\gamma^2}{4}$ \\
$\frac{\sigma\gamma}{2}$   & $\frac{{\sigma}^2\gamma^2}{4}$
\end{tabular}
\right)
\left(
\begin{tabular}{ll}
$\hat b_0$ \\
$\hat b_{00}$
\end{tabular}
\right)\,.
\eqa
The determinant of the matrix above is always different form zero,
for non vanishing $\lambda$ and $\sigma$,
when $\sigma \ne \lambda$, so that numerical inaccuracies
never occur.

Finally, the two last coefficients $\tld{b}_{01}$ and $\tld{b}_{02}$
are determined, in terms of
$q^+_{\lambda k}$ and $q^{\prime +}_{\sigma k}$, as solutions
of the system
\bqa
\left(
\begin{tabular}{l}
$Z (q^+_{\lambda k})$ \\\\
$Z (q^{\prime+}_{\sigma k})$
\end{tabular}
\right)
=
\left(
\begin{tabular}{ll}
$ -\lambda \gamma^2 F_\lambda e^{-i\pi/k}$
& $- \lambda \gamma^2 e^{i\pi/k}$ \\\\
  $- \sigma \gamma^2 e^{i\pi/k}$   & $-\sigma \gamma^2 F_\sigma e^{-i\pi/k}$
\end{tabular}
\right)
\left(
\begin{tabular}{ll}
$\tld{b}_{01}$ \\\\
$\tld{b}_{02}$
\end{tabular}
\right)\,,
\eqa
where
\bqa
Z(q) \equiv S(q)
  -\hat b_0    [(q+p_0) \cdot v]
  -\hat b_{00} [(q+p_0) \cdot v]^2  \,.
\eqa
Once again one verifies that when, for example, $k= 3$
the system never becomes singular.

\end{document}